\documentstyle[aps,multicol,pre,epsf]{revtex}
\begin{document}
\draft
\title { Dynamic algorithm for parameter estimation and its applications } 
\author{ Anil Maybhate$^{1,2,}$\footnote{e-mail: nil@prl.ernet.in}and 
         R.E.Amritkar$^{1,}$\footnote{e-mail: amritkar@prl.ernet.in}}
\address{ $^1$Physical Research Laboratory, Navrangpura, Ahmedabad 380009, India} 
\address{ $^2$Department of Physics, University of Pune, Pune 411007 India}
\maketitle

\begin{abstract} We consider a dynamic method, based on synchronization
and adaptive control, to estimate unknown parameters of a nonlinear
dynamical system from a given scalar chaotic time series. We present an
important extension of the method when time series of a scalar function of
the variables of the underlying dynamical system is given. We find that it
is possible to obtain synchronization as well as parameter estimation
using such a time series. We then consider a general quadratic flow in
three dimensions and discuss applicability of our method of parameter
estimation in this case. In practical situations one expects only a finite
time series of a system variable to be known. We show that the finite time
series can be repeatedly used to estimate unknown parameters with an
accuracy which improves and then saturates to a constant value with
repeated use of the time series. Finally we suggest an important
application of the parameter estimation method. We propose that the method
can be used to confirm the correctness of a trial function modeling an
external unknown perturbation to a known system. We show that our method
produces exact synchronization with the given time series only when the
trial function has a form identical to that of the perturbation.
\end{abstract}

\begin{multicols}{2}

\section{introduction}

An experimental observation often consists of reading a time series output
from a dynamical system. Such a time series can contain information about
the number as well as the form of the functions governing the evolution of
the system variables including nonlinearities (if any) and the parameters
\cite{ABST}. The estimation of parameter values from a given chaotic
scalar time series of a nonlinear system is the topic of our interest
here.

We have recently given a method to dynamically estimate unknown parameters
from the chaotic time series of a single phase space variable when the
system equations are known~\cite{MA}.  The method is based on a
combination of synchronization~\cite{PC1,PC2,CP} and adaptive
control~\cite{HL} similar to that used by John and
Amritkar~\cite{JA1,JA2}.

The problem of parameter estimation in nonlinear dynamics has been
considered earlier. Parlitz, Junge and Kocarev have given a static
method~\cite{PJK} based on minimization while Parlitz has developed a
method based on auto-synchronization~\cite{Par}. Unlike our method,
auto-synchronization method requires an ansatz for the parameter control
loop and gives slower convergences in many cases. A method requiring a
vector time series is given by Baker, Gollub and Blackburn~\cite{BGB} and
another method based on symbolic dynamics is discussed in
Refs.~\cite{TTBBB,TTB,Daw}. The effect of noise on parameter estimation
was studied by us~\cite{MA} and recently by Goodwin, Brown and
Junge~\cite{GBJ}. In contrast to many of these methods our method in
Ref.~\cite{MA} works asymptotically so that an exact estimation of the
parameters is in principle possible. The static methods based on
minimization are computationally expensive because they take a longer time
to run due to many iterations required for convergence and they also
require annealing to eliminate the possibility of getting trapped into a
local minimum.  The dynamic method as described in Ref.~\cite{MA} requires
only one time evolution of the system equations. The method also takes
care of annealing in a dynamic way.

In the first part of this paper we review our method for parameter
estimation in brief. We then extend it to a case when the time series of a
{\it scalar function} of phase space variables is given. We then go on to
study the applicability of the method to a general quadratic flow in three
dimensions.  This system has a large number of parameters and we try to
estimate some of them using our method.

In the second part, we show that it is possible to extend our method to a
more realistic situation, when the given time series is truncated after a
finite time. We find that a repetitive use of the finite time series can
be made to estimate the unknown parameters of the underlying system
without altering the dynamic nature of the method. The accuracy of such an
estimation increases with the increasing length of the given time series.
We also see that the accuracy saturates with the number of times the
finite time series is used.

Lastly in the third part of this paper, we suggest an interesting
application of parameter estimation method. Consider a situation where an
unknown perturbation disturbs a known chaotic system. In many practical
situations when the external perturbation is unknown, an ansatz function
modeling the behaviour of the external perturbation is tried. We show that
it is possible to use our parameter estimation method, to confirm the form
of an ansatz function modeling the external perturbation.

In section IIA we briefly introduce our method of parameter estimation and
discuss its important features. In section IIB we extend it to a general
situation when the given time series is obtained as a scalar function of
the phase space variables. Section IIC deals with a general quadratic flow
in three dimensions. In section III we extend the method to the case of
finite time series and present two examples. Finally in section IV we give
the application of the method in confirming the form of an unknown
external perturbation to a known dynamical system.  In section V we
conclude with a summary of the results.

\section{parameter estimation} 

\subsection{The method}

Here, we briefly introduce our method for parameter estimation from a
scalar time series. We would like to direct the reader to Ref.~\cite{MA}
for a more detailed discussion. We start by considering an autonomous
dynamical system of the form,
\begin{eqnarray}
{\bf\dot x} = {\bf f}\left({\bf x},\alpha\right),
\label{TRANS}
\end{eqnarray}
where ${\bf x} = (x_1, x_2, \ldots, x_n)$ is an $n$-dimensional state
vector whose evolution is described by the function ${\bf f}=(f_1, \ldots,
f_n)$. We denote a set of $m$ unknown scalar parameters by
$\alpha=(\alpha_1, \alpha_2, \dots, \alpha_m) $. A possible appearance of
any other parameters (assumed to be known) is not shown in
Eq.~(\ref{TRANS}).

Without loss of generality we assume that a time series of the variable
$x_1$ is given. The problem we consider is to estimate $\alpha$ from the
given scalar time series of $x_1$ assuming the functional form of {\bf f}
to be known.

In analogy with the control method used earlier by John and
Amritkar~\cite{JA1,JA2}, we combine synchronization with adaptive control
to achieve our goal of estimating $\alpha$ in Eq.~(\ref{TRANS}) as
follows.  We construct another system of variables ${\bf x'}$ having a
structure identical to that of Eq.~(\ref{TRANS}) with a linear feedback
proportional to the difference $x'_1-x_1$ added in the evolution of the
variable $x_1$. Thus the system is given by,
\begin{eqnarray}
\dot x'_1 &=& f_1({\bf x'},\alpha') - \epsilon (x'_1-x_1) \nonumber \\
\dot x'_j  &=& f_j ({\bf x'},\alpha'), \;\;\; j=2,\dots,n.
\label{RECEV}
\end{eqnarray}
where the function ${\bf f}=(f_1, \ldots, f_n)$ is the same as that in
Eq.~(\ref{TRANS}). The initial values of parameters $\alpha'$ which
correspond to the unknown parameters $\alpha$ in Eq.~(\ref{TRANS}) are
chosen randomly. The newly introduced parameter $\epsilon$ is the feedback
constant. It is known that if $\alpha'=\alpha$ then the systems
(\ref{TRANS}) and (\ref{RECEV}) synchronize after an initial transient,
provided the conditional Lyapunov exponents (CLE's) of the system
(\ref{RECEV}) are all negative~\cite{MA}. The CLE's are obtained from the
eigenvalues of the Jacobian matrix $J$ whose elements are given by,
\begin{equation}
J_{ij} = {\partial f_i \over \partial x_j} - \epsilon \delta_{i1}
\delta_{j1}
\label{JAC}
\end{equation}

Since the values $\alpha=(\alpha_1, \dots, \alpha_m)$ are unknown, we need
to set $\alpha'=(\alpha'_1, \dots, \alpha'_m)$ to random initial values
and evolve them {\it adaptively} so that they converge to the values
$\alpha$. Note that a good guess for the initial values of $\alpha'$, can
be useful in many cases.

We first consider the case when $\alpha$ (and its counterpart $\alpha'$)
contains only a single element, i.e. the case when only a single parameter
in Eq.~(\ref{TRANS}) is unknown. For notational simplicity we now denote
this single parameter by $\alpha$. We start with a random initial value
for $\alpha'$ and evolve it in a controlled fashion so that it converges
to $\alpha$. This is achieved by raising $\alpha'$ to the status of a
variable which evolves as,
\begin{equation}
\dot\alpha'=-\delta(x'_1-x_1)\;w\left({\partial
f_1\over\partial\alpha'}\right),
\label{CNTRL}
\end{equation}
where $\delta$ is called stiffness constant and $w$ is some suitably
chosen function of ${\partial f_1 /\partial \alpha'}$. A simple choice for
$w$ is $ w = {\partial f_1 / \partial \alpha'}$ giving the adaptive
evolution equation for $\alpha'$ as,
\begin{equation}
\dot\alpha'=-\delta(x'_1-x_1){\partial f_1\over\partial\alpha'}.
\label{ADAPT}
\end{equation}
  
Eq.~(\ref{CNTRL}) or Eq.~(\ref{ADAPT}) when coupled with Eq.~(\ref{RECEV})
constitutes our method of parameter estimation. A vector $({\bf x'},
\alpha')$ initially set to random values asymptotically converges to a
vector $({\bf x}, \alpha)$ in Eq.~(\ref{TRANS}) provided the conditional
Lyapunov exponents (CLE's) for the combined system (Eqs.~(\ref{RECEV}) and
(\ref{ADAPT})) are all negative. This facilitates the estimation of
$\alpha$.

Eq.~(\ref{ADAPT})  is equivalent to a dynamic algorithm for minimization
of synchronization error between Eqs.~(\ref{TRANS}) and (\ref{RECEV}) as
discussed in Ref.~\cite{MA}.

Note that if we assume in the above discussion that the unknown parameter
$\alpha$ appears in the function $f_1$ corresponding to the variable $x_1$
for which the time series is given then the calculation of the factor
$\partial f_1 /\partial \alpha'$ in Eq.~(\ref{ADAPT}) is straightforward.
However this may not be necessarily the case. The parameter $\alpha$ may
appear in any of the other system functions. If it appears in the
functions for the variables for which the time series is not given, e.g.
in any of the functions $f_2, \dots, f_n$ in Eq.~(\ref{TRANS}), then
correspondingly the calculation of the factor $\partial f_1 /\partial
\alpha'$ becomes nontrivial.

To make this point clear we assume that the unknown parameter $\alpha$
appears in the function $f_k ({\bf x})$ governing the evolution of
variable $x_k$ with $k\neq 1$ while the time series of $x_1$ is given.  
In such a case Eq.~(\ref{ADAPT}) gets modified to, (See Ref.~\cite{MA})
\begin{equation} 
\dot\alpha'=-\delta(x'_1-x_1){\partial f_1\over\partial
x'_k}{\partial f_k\over\partial\alpha'}. 
\label{ADAPT2} 
\end{equation}

Further if the variable $x_k$ itself does not appear in the function $f_1$
then the complexity of the calculation still increases. This issue has
been explained in detail with an example in Ref.~\cite{MA}.

Next we consider the case when the set $\alpha$ of unknown parameters
contains more than one element, say $(\alpha_1, \alpha_2, \dots)$. Now we
set up an adaptive evolution for each of the corresponding parameters
$(\alpha'_1, \alpha'2, \dots)$. For the case of two unknown parameters
$\alpha_1$ and $\alpha_2$, appearing in functions $f_k$ and $f_l$
respectively, the adaptive evolution is given by,
\begin{eqnarray}
\dot\alpha'_1&=&-\delta_1(x'_1-x_1){\partial f_1\over\partial x'_k}
{\partial f_k\over\partial\alpha'} \nonumber\\
\dot\alpha'_2&=&-\delta_2(x'_1-x_1){\partial f_1\over\partial x'_l}
{\partial f_l\over\partial\alpha'}, 
\label{ADAPTWO}
\end{eqnarray} 
where $\delta_1$ and $\delta_2$ are two stiffness constants deciding the
rates of convergence. For estimating the values of $\alpha_1$ and
$\alpha_2$ Eqs.~(\ref{ADAPTWO}) can be coupled with Eqs.~(\ref{RECEV})
which provide the necessary synchronization of system variables if the
associated CLE's are negative.

In the next subsection we extend our method to a situation when a time
series of a {\it scalar function} of phase space variables is given. We
show that it is not only possible to build a synchronizing system but also
to adaptively estimate an unknown parameter.

\subsection{Parameter estimation using time series of a scalar function of
variables}

In our discussion of parameter estimation in earlier subsection, we have
assumed that time series of one of the phase space variables is given.
This may not be the case in many practical applications and in general the
observed quantity can be a function of the phase space variables, say
$s({\bf x})$. It is possible to construct a synchronization scheme in such
a situation~\cite{MA1}.

We consider the system given by Eq.~(\ref{TRANS}) and assume that the time
series $s({\bf x})$ which is a function of phase space variables is given.
A synchronization scheme can be set up in this case by using a suitable
modification of the feedback in Eq.~(\ref{RECEV}) as follows~\cite{MA1}.
\begin{eqnarray}
\dot x'_1&=& f_1({\bf x'},\alpha)-\epsilon\;\rm{sgn}
\left(\partial s' \over \partial x'_1\right)(s'-s({\bf x}))
\nonumber\\ 
\dot x'_ j &=& f_ j({\bf x'},\alpha) \;\;\;  j=2,\dots,n.
\label{SYNF}
\end{eqnarray}
where $s'=s({\bf x'})$ and we give a feedback proportional to $(s'-s)$ in
the function $f_1$ with feedback constant $\epsilon$. The function $s({\bf
x})$ denotes the given time series.

It can be shown that if the parameters $\alpha$ are assumed to be known,
the above system of equations for $x'$ (Eqs.(\ref{SYNF})) converges to
$x$, provided the CLE's are all negative~\cite{MA1}.

In Eqs.~(\ref{SYNF}), we have assumed that $s({\bf x})$ has an explicit
dependence on the variable $x_1$ so that ${\partial s'/\partial x'_1} \neq
0$. If this is not the case, we can choose any other variable for the
feedback on which $s({\bf x})$ depends explicitly. The factor $\rm{sgn}()$
in Eq.~(\ref{SYNF})  makes sure that the term provides a `negative
feedback' for all the time so that a convergence is feasible.

To estimate parameter $\alpha$ in such a case, we set up a synchronization
scheme combined with an adaptive control in analogy with
Eqs.~(\ref{RECEV}) and ~(\ref{CNTRL}). This system can be written as,
\begin{eqnarray}
\dot x'_1&=& f_1({\bf x'},\alpha')-\epsilon\;\rm{sgn}
\left(\partial s' \over \partial x'_1\right)(s'-s({\bf x}))
\nonumber\\ 
\dot x'_ j &=& f_ j({\bf x'},\alpha') \;\;\;  j=2,\dots,n.
\nonumber\\
\dot \alpha'&=&-\delta\;{\rm sgn}\left({\partial s' \over\partial x'_1}\right)
(s'-s({\bf x})) {\partial f_1 \over\partial \alpha'}.
\label{RECEF}
\end{eqnarray}
Eqs.~(\ref{RECEF}) can be used for estimating $\alpha$ when a time series
of $s({\bf x})$ is given. The condition for such an estimation of $\alpha$
to be possible is that the CLE's associated with the system~(\ref{RECEF})
are all negative.

To demonstrate the above procedure, we consider the Lorenz system given by,
\begin{eqnarray}
\dot x &=& \sigma (y-x)\nonumber\\
\dot y &=& rx-y-xz \nonumber\\
\dot z &=& xy-bz,
\label{LORTF}
\end{eqnarray}
where the variables $(x, y, z)$ define the state of the system while
$(\sigma, r, b)$ are the three parameters. We consider the case when the
time series of $s(x,y,z)=0.5x^2+1.1y$ is given as an output of the above
system and the parameter $\sigma$ is unknown.

To estimate the value of $\sigma$, we form a system of variables $(x', y',
z', \sigma')$ similar to Eq.~(\ref{RECEF}). The evolution equations are
\begin{eqnarray}
\dot x' &=& \sigma'(y'-x') - \epsilon\; {\rm sgn}(x') (s' - s(x,y,z)) \nonumber \\
\dot y' &=& rx'-y'-x'z' \nonumber\\
\dot z' &=& x'y'-bz'\nonumber\\
\dot \sigma' &=& -\delta\;{\rm sgn} (x') (s'-s(x,y,z)) (y'-x'), 
\label{LORRF}
\end{eqnarray}
where $s' =0.5x'^2+1.1y'$.

Figures 1(a)-(d) show the evolution of the differences $x'-x, y'-y, z'-z,
\sigma'-\sigma$ respectively (Eqs.~(\ref{LORTF}) and~(\ref{LORRF})) as a
function of time $t$. We see that these differences all go to zero as $t
\rightarrow \infty$.  This indicates that an unknown $\sigma$ can be
estimated using Eq.~(\ref{LORRF}).

The CLE's are obtained using the Jacobian matrix $J$ given by
\begin{equation}
J = \left( \begin{array}{cccc}
-\sigma - \epsilon \, {\rm sgn}(x) x & \sigma - 1.1 \epsilon \, {\rm sgn}(x) 
& 0 & y-x \\
r-z & -1 & -x & 0 \\
y & x & -b & 0 \\
-\delta \, {\rm sgn}(x) x (y-x) & - 1.1 \delta \, {\rm sgn}(x) (y-x) & 0 & 0
\end{array} \right)
\label{CLELOR}
\end{equation}
We have verified that all the CLE's are less than zero except one trivial
CLE which is zero.

We have performed simulations and successfully estimated unknown
parameters in Lorenz system with other forms of the function $s(x, y, z)$.
The function $s(x, y, z)$ should however be such that all the associated
conditional Lyapunov exponents should be negative.

\subsection{A general quadratic flow in 3-D}

Now  we consider a quadratic flow in 3-D given by,
\begin{eqnarray}
\dot x = &  & a_0 + a_1 x + a_2 y + a_3 z + a_4 x^2 + a_5 y^2 \nonumber\\ 
             &+& a_6 z^2 + a_7 xy + a_8 yz + a_9 xz \nonumber\\
\dot y = &  & b_0 + b_1 x + b_2 y + b_3 z + b_4 x^2 + b_5 y^2 \nonumber\\ 
             &+& b_6 z^2 + b_7 xy + b_8 yz + b_9 xz \nonumber\\
\dot z = &  & c_0 + c_1 x + c_2 y + c_3 z + c_4 x^2 + c_5 y^2 \nonumber\\ 
             &+& c_6 z^2 + c_7 xy + c_8 yz + c_9 xz, 
\label{QUAD} 
\end{eqnarray}
where $(a_0, \dots, a_9, b_0, \dots, b_9, c_0, \dots, c_9)$ form a thirty
dimensional parameter space and $(x, y, z)$ are the three variables. We
have performed simulations in which we have assumed more than one of the
thirty parameters of the system (\ref{QUAD}) to be unknown and tried to
estimate them when a time series of one of the variables is given.

To elaborate, we assume some of the thirty parameters to be unknown while
the remaining to be known.  Some of the known or unknown parameters may be
zero thereby making the corresponding term absent from the system. To
illustrate the procedure we consider a case when three parameters $(a_1,
a_2, a_7)$ are unknown and a time series of $x$ is given, we set up a
system of equations similar to Eq.~(\ref{RECEV}) with the adaptive control
loops similar to Eq.~(\ref{ADAPTWO}) for the three parameters $(a'_1,
a'_2, a'_7)$ as,
\begin{eqnarray}
\dot a'_1 &=& -\delta_1 (x'-x) x' \nonumber \\
\dot a'_2 &=& -\delta_2(x'-x) y'               \nonumber \\
\dot a'_7 &=& -\delta_3(x'-x) x'y'.
\label{ADAPQ}
\end{eqnarray}

Eqs.~(\ref{ADAPQ}) when coupled to the system of variables $(x', y', z')$
with an identical structure of evolution as Eq.~(\ref{QUAD}) with a
feedback term in the evolution of $x'$, can provide the necessary
estimation of parameters when the CLE's associated with the reconstructed
system are all negative.

In Fig.2(a)-(c) we plot the time evolution of the differences $a'_1-a_1,
a'_2-a_2, a'_7-a_7$ as a function of time. The correct value of $a_7$ was
zero while the other two were non-zero. All the differences go to zero
indicating the feasibility of simultaneous estimation of the three
parameters $(a_1, a_2, a_7)$ even when the actual value of one of them is
zero. This shows that the method does not falsely detect a term which is
absent in the system.

We have found cases when our method can be used successfully for the
system (\ref{QUAD}) to simultaneously estimate as many as five parameters.
(One such case is the set of parameters $a_1, a_2, a_7, b_3, c_1$, while
the time series of $x$ is given.)

Further we have also found that when {\it any} two of the thirty
parameters in the system~(\ref{QUAD}) are unknown, we can apply our method
to simultaneously estimate them asymptotically {\it to any desired
accuracy} when the time series of a suitably chosen variables is given.
Our results suggest that the information about all the thirty parameters
should in principle be contained in the time series of a single variable
of the system, though at present we do not have any systematic approach to
the simultaneous estimation of all of them.

\section{parameter estimation using a finite time series}

\subsection{Algorithm for repetitive use}

In this section we discuss an algorithm for repetitive use of our method
to impove the accuracy of parameter estimation when the given time series
is of finite duration.

Before going on to describe the algorithm it should be mentioned here that
even if a finite time series is used repeatedly, we do not expect an exact
estimation of the unknown parameter. A finite chaotic trajectory sets a
limit on the accuracy to which the unknown parameter can be estimated.
This can be seen as follows :

We consider symbolic dynamics on the attractor which provides
a generating partion of the attractor. It is well known that as the system
evolves in time, a finer and finer coarse graining is required to specify
a particular trajectory or alternatively, the trajectory gives us a finer
coarse grained information about the attractor. The number of coarse
grained partitions as a function of time goes as,
\begin{equation}
n_p \sim exp\{ht\},
\label{np1}
\end{equation}
where $h$ is the Kolmogorov entropy~\cite{Nic}. 

If $\xi^d$ is the volume of a hypercube in a $d$ dimensional phase space
and if the size of the attractor is normailized to unity, the number of
hypercubes in a generating partion may be approximated as,
\begin{equation}
n_p \sim {1\over{\xi^d}}.
\label{np2}
\end{equation}
Equations (\ref{np1}) and (\ref{np2}) indicate that the length scale of a
hypercube in a generating partition goes as,
\begin{equation}
\xi \sim exp\{-{h \over d}t\}.
\label{xi}
\end{equation}
It can be seen from Eq.~(\ref{xi}) that as long as $t$ is finite, the
volume of the hypercube in a coarse graining of the attractor will not
reduce to zero. Thus a finite trajectory sets a limit on the accuracy to
which any information can be extracted from it. This can be further
related to Lyapunov exponents using the famous Kaplan-Yorke
conjecture~\cite{KY} as,
\begin{equation}
\xi \sim exp\{-{{\sum_{\lambda>0} \lambda } \over d}t\},
\label{KY}
\end{equation}
where $\lambda$ is the characteristic Lyapunov exponent of the system. For
a chaotic system with a single positive Lyapunov exponent denoted by
$\lambda^+$, eqn.~(\ref{KY}) reduces to,
\begin{equation}
\xi \sim exp\{-{{\lambda^+} \over d}t\}.
\label{KY+}
\end{equation}

Now we will discuss the algorithm for repetitive use of a finite time
series to estimate an unknown parameter. Similar to the case considered in
section IIA, we assume that the parameters $\alpha = (\alpha_1, \alpha_2,
\dots, \alpha_m)$ in Eq.(\ref{TRANS}) are unknown while the time series of
$x_1$ is given. We further assume that the time series is truncated after
a finite time $T$.

For the time interval $0\leq t\leq T$, we can use the procedure identical
to that described earlier (Eqs.~(\ref{RECEV}) and~(\ref{ADAPT})) to evolve
variables $({\bf x'},\alpha')$ with random initial conditions. The given
finite time series is fed in system (\ref{RECEV}) as in the earlier case.  
In this way we can get an approximate value of $\alpha$ which we denote
as, $\alpha^1=\alpha'(T)$.

Now at time $t=T$ we set the variables ${\bf x'}$ to exactly the same
(randomly chosen earlier) initial values while $\alpha' = \alpha^1$ and
feed the same finite time series $\{x(t)|0\leq t \leq T\}$ again into the
system (\ref{RECEV}) through the feedback terms in Eqs.~(\ref{RECEV})
and~(\ref{ADAPT}), i.e. we set $x(t+T)=x(t)$. We now evolve the variables
$({\bf x'}, \alpha')$ for the time interval $T\leq t \leq 2T$ to obtain a
new estimated value of $\alpha$ which is $\alpha^2 = \alpha'(2T)$.

We repeat the procedure to get successive estimates for the value of
$\alpha$ denoted by $\alpha^1, \alpha^2, \alpha^3, \dots, \alpha^N, \dots$
at times $t=T,2T, 3T, \dots, NT, \dots$ respectively. Thus, starting from
an initial guess for the value of $\alpha$ we obtain a sequence of
estimates $\alpha^0, \alpha^1, \dots, \alpha^N$ after $N$ usages of the
given finite time series. For large enough $N$ we get a better and better
estimate of $\alpha$, although eventually the accuracy of such an estimate
saturates as $N$ is increased further.

The conditions for the method of parameter estimation using a finite time
series to work successfully are~:

\begin{enumerate} \item{The conditional Lyapunov exponents associated with
the reconstructed system should be all negative.} \item{The time $T$ after
which the given time series is truncated should satisfy $T>\tau$ where
$\tau$ denotes the transient time required for synchronization of the
systems (\ref{TRANS}) and (\ref{RECEV})  with the parameter evolution
given by Eq.~(\ref{ADAPT}).} \end{enumerate}

In the next subsection, we discuss two examples of parameter estimation
from a finite time series, viz. Lorenz system and an electrical circuit of
a phase converter.
 
\subsection{Examples}

\subsubsection{Lorenz system}

As our first example we choose the Lorenz system given by
Eq.~(\ref{LORTF}) where we assume that the time series $\{x(t)|0\leq t
\leq T\}$ is given and the value of $\sigma$ is to be estimated.  We set
up the following system of equations (see Eqs.~(\ref{RECEV})
and~(\ref{ADAPT})).
\begin{eqnarray}
\dot x' &=& \sigma(y'-x') -\epsilon(x'-x) \nonumber\\
\dot y' &=& rx'-y'-x'z'\nonumber\\
\dot z' &=& x'y'-bz' \nonumber \\
\dot \sigma' &=& -\delta (x'-x) (y'-x')
\label{LORR}
\end{eqnarray}
where we feed the given time series in the evolution of $x$ for the
interval $0\leq t \leq T$ to obtain the first estimate $\sigma^1$.

As described in the earlier subsection, we then go on repetitively feeding
the same finite time series $x(t)$ in Eq.(\ref{LORR}) to obtain successive
estimates for the value of $\sigma$.  Starting from a random initial value
we denote this sequence of estimates by $\sigma^0,\sigma^1,\dots,\sigma^N$
where N denotes the number of times we use the given time series.

In Fig.3 we plot the evolution of the difference $\sigma'-\sigma$ as a
function of time $t$ during the time interval $0\leq t \leq 3T$ where we
use the time series $x(t)$ thrice. We see that the difference decreases as
we increase the number of times the finite time series is used.  We also
observe that shortly after each resetting of the initial vector $(x', y',
z')$ which is done at times $T, 2T$, the synchronization weakens and
fluctuations are present. This is due to the random resetting of the $y$
and $ z$ components which gives a transient before the synchronization is
recovered.  An appropriate feedback constant $\epsilon$ may be chosen to
lessen this transient in every usage of the time series.

In Fig.4 we plot the successive differences $\sigma^N-\sigma$ as a
function of $N$, the number of times we use the given finite time series.  
We see that the difference $\sigma^N-\sigma$ goes on decreasing with
increasing $N$. However, as $N$ is increased further, it saturates to a constant
finite value depending on the length of the time series used for the
calculations. This is consistent with our expectations that finite time series
can contain only finite information about the system as discussed in the
previous subsection, e.g. using $\lambda^+ \sim 0.9$,
the finest length scale that can be obtained using a finite time series
with $T=30$ is estimated to be $0.05$ (Eq.~(\ref{KY+})) which means an
accuracy of about $10^{-3}$. This is also the order of magnitude of the
accuracy of parameter estimation.

The three curves in Fig. 4 correspond to three different values of $T=
T_1<T_2<T_3$. We see that an increasing $T$ gives better estimate of the
parameter. This is natural since a very long time series corresponding to
$T \rightarrow \infty$ is expected to give an exact estimation of the
unknown parameter.

We have similarly implemented our method to estimate other parameters of
Lorenz system using finite time series of either $x$ or $y$. The method
fails to estimate any of the parameters when the time series of $z$ is
given. The reason for this is that one of the associated conditional
Lyapunov exponents is critically zero and the convergences are slow.

\subsubsection{A phase converter circuit}

As our next example, we consider the set of equations describing an
electrical circuit for a phase converter~\cite{YK} system in a
dimensionless form given by,
\begin{eqnarray}
\dot x_1 &=&  x_2\nonumber \\
\dot x_2 &=&  -kx_2-{x_1\over 4}\left(x_1^2+3x_3^2\right) \nonumber \\
\dot x_3 &=&  x_4\nonumber \\
\dot x_4 &=&  -kx_4-{x_3\over 4}\left(x_1^2+3x_3^2\right) + B \cos  t 
\label{CKT}
\end{eqnarray}
where $k$ and $B$ are the two parameters. Here we consider the time series
$\{x_2(t)|0\leq t \leq T \}$ to be given. Notice that the
system~(\ref{CKT}) has a simple time dependent term making it a
non-autonomous system. Such a system is equivalent to an autonomous system
in higher dimensions. We have successfully estimated any one of the
parameters $k$ or $B$ (or both) using finite time series of $x_2(t)$.

Figure 5(a) shows a schematic diagram of the circuit for the phase
converter. The system in known to exhibit a chaotic behaviour due to
period doubling bifurcations, codimension two bifurcations etc. Figure 5
(b) shows a chaotic attractor in the $x_1-x_2$ plane of the phase space.

Figure 6 shows the plot of the successive differences $k^N-k$ as a
function of $N$, the number of times we use the given time series for two
different values of the truncation time T. As expected the accuracy of the
estimation increases with increasing $T$ while showing a saturation with
increasing number of repeated usages.

Thus, we have shown how the method of parameter estimation can be used
when a finite time series is given. The method works when the CLE's
associated are all negative and the time series given is of longer
duration than the transient time required for synchronization.

\section{Form of a model perturbation}

Here we describe an interesting application of our method to test a
function modeling an unknown external source of perturbation to a known
chaotic system. In many practical situations when an external source of
disturbance in not known, a trial function is used to model the
perturbation.

We imagine a situation when it is required to verify a proposed trial
model form for the perturbation. We denote the actual perturbation by a
function $F({\bf x}, \mu)$ and the trial function by $G({\bf x'}, \mu')$
where $\mu$ and $\mu'$ are parameters.  In the following, we demonstrate
the use of our method of parameter estimation to confirm the form of the
trial function.  Note that here we do not deal with the issue of obtaining
the form of the model function.

Now if the proposed trial function $G$ models the external perturbation
$F$ correctly, then a scheme based on synchronization combined with
adaptive control should produce synchronization of variables and make the
parameters $\mu'$ converge (to $\mu$).  Thus a successful synchronization
should then indicate a correctly chosen model function. In this manner we
can use the method to distinguish between a correct model and a wrong
model for an external perturbation.  We elaborate on this application
further using the example of Lorenz system.

Consider the Lorenz system perturbed by a sinusoidal term $F=A\sin(\omega
x)$,
\begin{eqnarray}
\dot x &=& \sigma(y-x) + A \sin(\omega x) \nonumber \\
\dot y &=& r x-y-x z \nonumber \\
\dot z &=& x y - b z, 
\label{LOREX}
\end{eqnarray}
where we assume the unperturbed Lorenz system to be known. The function
$F=A \sin(\omega x)$ is the external perturbation.  We assume that the
time series of $x$ is given as an output of the system~(\ref{LOREX}).

To set up the required scheme we construct a system of variables $(x', y',
z')$ and their evolution as,
\begin{eqnarray}
\dot x' &=& \sigma(y'-x') + G(x',y',z',\mu') -\epsilon(x'-x) \nonumber\\
\dot y' &=& r x'-y'-x' z' \nonumber \\
\dot z' &=& x' y' - b z', \nonumber\\
\dot \mu' &=& -\delta (x'-x) {\partial G \over \partial \mu'}. 
\label{MODEL}
\end{eqnarray}
where $G(x', y' z')$ is the trial perturbation function.

We feed the time series $x(t)$ obtained from system~(\ref{LOREX})  into
the model system~(\ref{MODEL}). Now if $G$ models the behaviour of $F$
correctly then the two systems should exhibit synchronization while the
parameters should show convergence to the correct values. In our
simulations we have tried several different forms for the trial function
$G$.

Figures 7(a)-(c) show the time evolution of $x'-x, \mu_1$ and $\mu_2$
respectively while the feedback is given into $x$ and the trial function
is $G=\mu_1 x^2 + \mu_2$. It can be clearly seen that there is no
synchronization of variables. The trial function $G = \mu_1 x^2 + \mu_2$
thus fails to produce synchronization and hence can be discarded as a
plausible model for $F$. We also note that the parameters $\mu'_1$ and
$\mu'_2$ do not show convergence.

In Figs. 8 and 9, we plot similar graphs for two more choices of the trial
function. In Fig. 8(a)-(c) we use $G=\mu_1 x-\mu_2 x^3$ and plot $x'-x,
\mu_1$ and $\mu_2$ respectively.  We choose this form of $G$ since it
represents the two leading terms in the series expansion of the the
function $F=A\sin(\omega x)$.  We can see from Fig. 8 that such an
approximation fails to produce synchronization and also the convergence of
parameters.

As a third choice we use $G=\mu_1 \sin(\mu_2 x)$ in Eq.~(\ref{MODEL}) and
plot the time evolution of $x'-x, \mu_1$ and $\mu_2$ in Fig.9(a)-(c)
respectively. The difference $x'-x$ goes to zero as time increases showing
synchronization. The parameters $\mu_1$ and $\mu_2$ converge to the
correct values $A$ and $\omega$ respectively. The variables $y'$ and $z'$
also synchronize with $y$ and $z$ respectively. This confirms that this
trial function correctly models the function $F$.

Now as a last consideration, we use the form $G=\mu_1 \sin(\mu_2 x)$ again
but unlike in Eq.~(\ref{MODEL}) we perturb a wrong variable in the model
system, i.e. we choose to add the trial perturbation in the evolution of
say $y'$.  The feedback is given in $x$. The evolution equations are
\begin{eqnarray}
\dot x' &=& \sigma(y'-x')  -\epsilon(x'-x) \nonumber\\
\dot y' &=& r x'-y'-x' z' + G(x',y',z',\mu') \nonumber \\
\dot z' &=& x' y' - b z', \nonumber\\
\dot \mu' &=& -\delta (x'-x) {\partial G \over \partial \mu'}. 
\label{MODELY}
\end{eqnarray}

In Fig.10~(a)-(c) we plot the time evolution of $x'-x, \mu_1$ and $\mu_2$
respectively.  We see that even if $G$ correctly models $F$,
synchronization does not take place. This shows that along with the form
of $F$ we can also confirm a guess about the perturbed variable.

Thus, the results presented in this section suggest that the method which
we use for estimating parameters can be used to distinguish between a
correct trial function and the wrong trial functions for an unknown
external perturbation to a known system~\cite{Note}.

\section{summary and conclusions}

We have described dynamic method of parameter estimation from a given
chaotic time series of a phase space variable of a dynamical
system~\cite{MA}.
Further, We have generalized the method for the case when the quantity for
which the time series is given is a {\it scalar function} of the phase
space variables. We have shown that it is not only possible to synchronize
two systems using the time series of the scalar function but also to
asymptotically estimate unknown parameters adaptively to any desired
accuracy. This is done by providing a linear feedback in the evolution of
one of the variables on which the scalar function explicitly depends. The
method works successfully provided the function for which the time series
is given is such that the associated conditional Lyapunov exponents are
all negative.

We have also applied our method to a system with a large number of
parameters, i.e. a general quadratic flow in 3-D.  We have observed that a
simultaneous estimation of a few parameters is possible provided the
condition of convergence as stated in Ref.~\cite{MA} is satisfied i.e. all
the CLE's are negative.     

As a next consideration, we have extended our method to a realistic
situation when the given series is truncated after a finite time. We have
shown that repetitive use of a finite time series can be made to estimate
an unknown parameter of the system. The accuracy of the parameter
estimation saturates as the given finite time series is used more and more
number of times. The accuracy increases with the increasing length of the
given time series.

In the end we have demonstrated an important application of our method in
confirming the correctness of a trial model function for an unknown
external perturbation to a known system. We see that a perfect
synchronization between a perturbed system and its dynamical copy using a
model for the perturbation is possible only when the form of the trial
function is correctly guessed. These results indicate that our method can
be used as a test for the trial model for an unknown external perturbation
to a known system. Another possible application (not discussed in the
paper) is as follows.
Our method may be
employed to experimentally measure the unknown value of a component
added to a known circuit. In such a situation the equations
governing the circuit are known, and can be used to
estimate the unknown component value accurately. This is feasible due to
the asymptotic convergences in our method.

\acknowledgements

One of the authors (A.M.) would like to thank U.G.C.(India) for financial
support.

\end{multicols}

\newpage 

\begin{figure} 
\epsffile{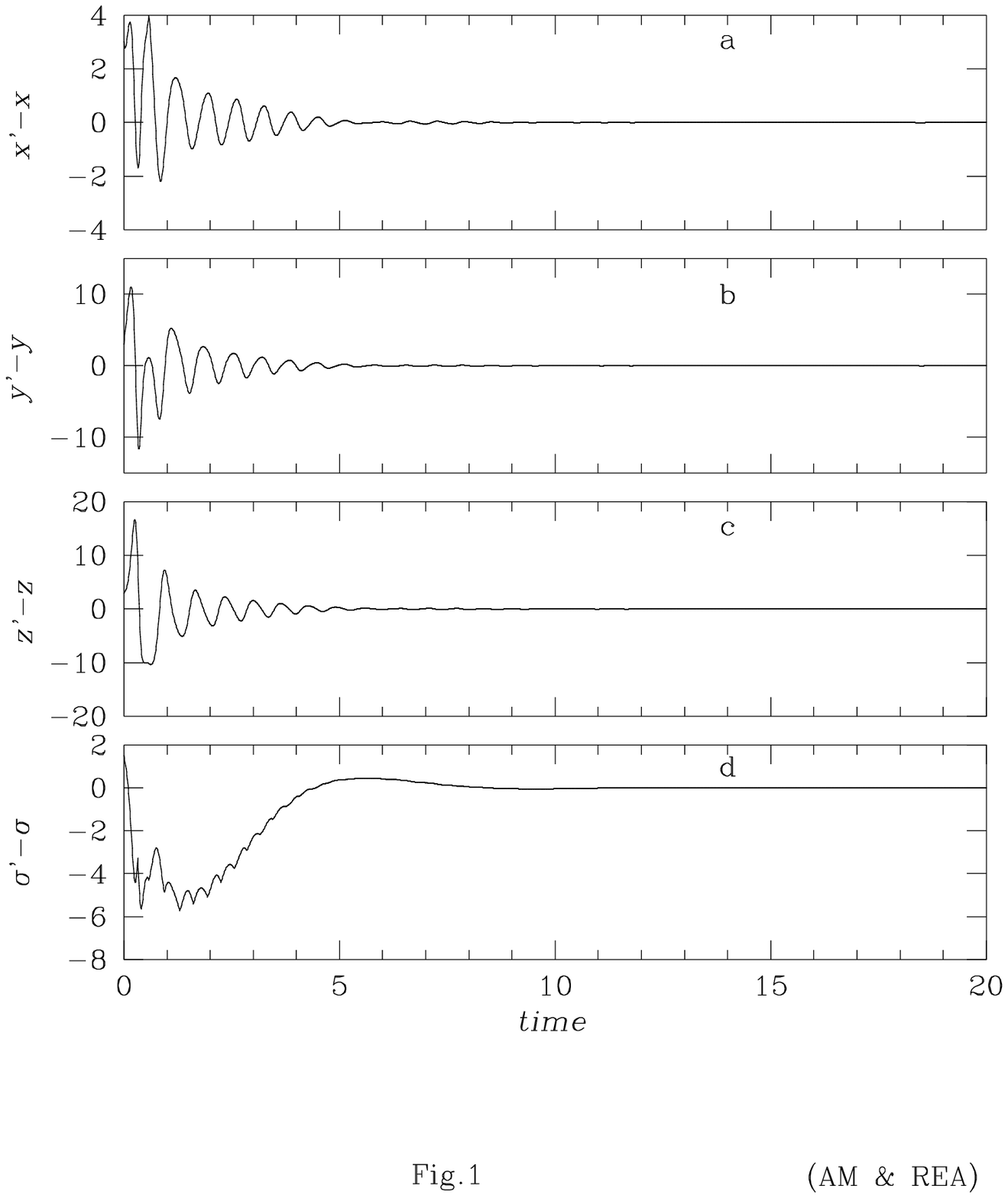}
\caption{ The plots (a)-(d) show the evolution of the
differences $x'-x, y'-y, z'-z,\sigma'-\sigma$ as a function of time for
the Lorenz system (Eqs.~(\ref{LORTF}) and ~(\ref{LORRF})) respectively for
the case when a time series for $s(x,y,z)=0.5x^2+1.1y$ is given and
$\sigma$ is unknown.  The differences go to zero asymptotically indicating
that it is possible to use our method to estimate an unknown parameter
when time series for $s({\bf x})$ is given.} \end{figure}

\begin{figure} 
\epsffile{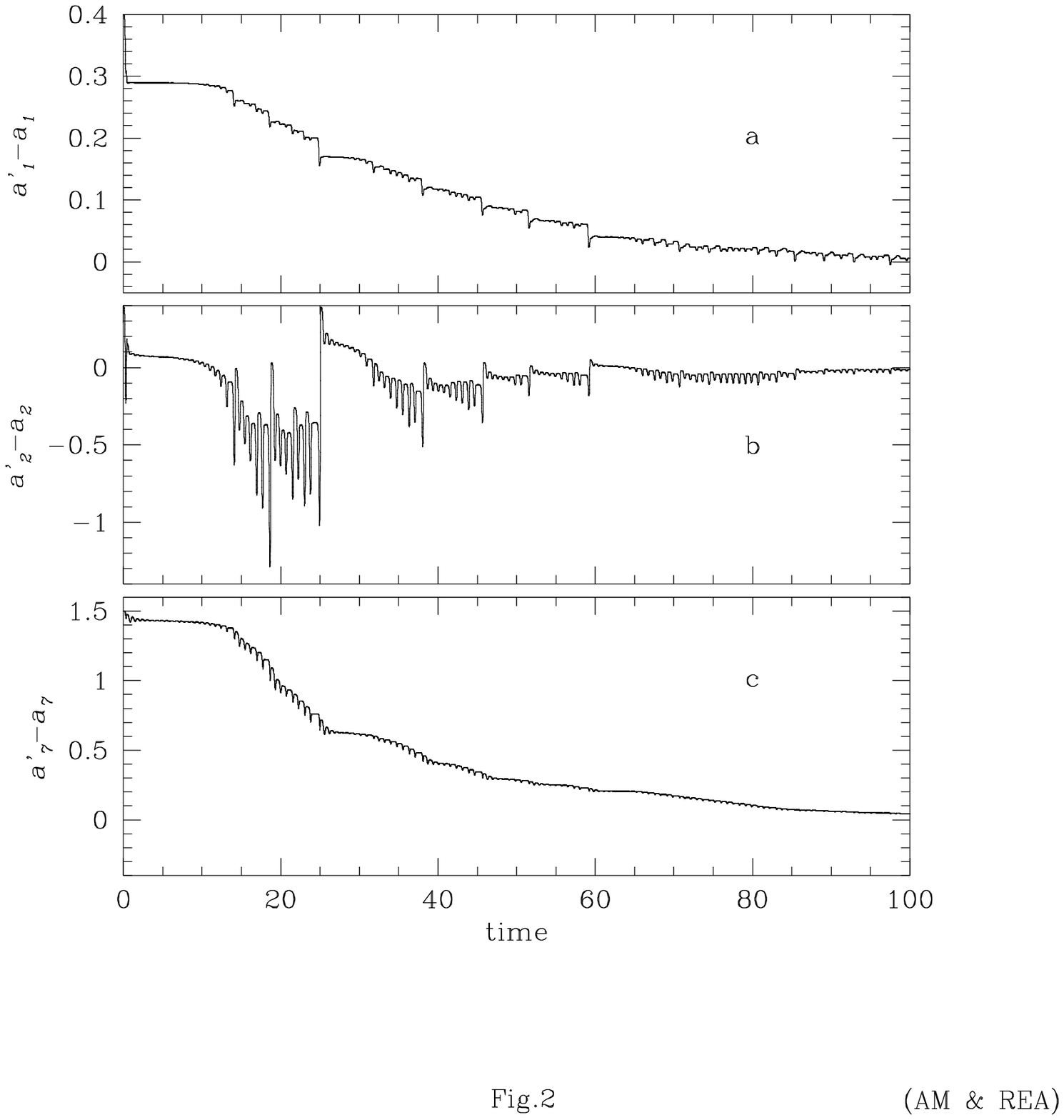}
\caption{ The plots (a)-(c) show the evolution of the
differences $a'_1-a_1, a'_2-a_2,a'_7-a_7$ for a general quadratic flow in
3-D (Eq.~(\ref{QUAD}) and~(\ref{ADAPQ})), plotted as a function of time
when the time series of $x$ is given. We see that all the differences
approach zero indicating the feasibility of simultaneous estimation of
more than one parameter. The correct value of $a_7$ was zero, showing that
a term absent in the flow equations is not falsely detected by our
method.} \end{figure}

\begin{figure} 
\epsffile{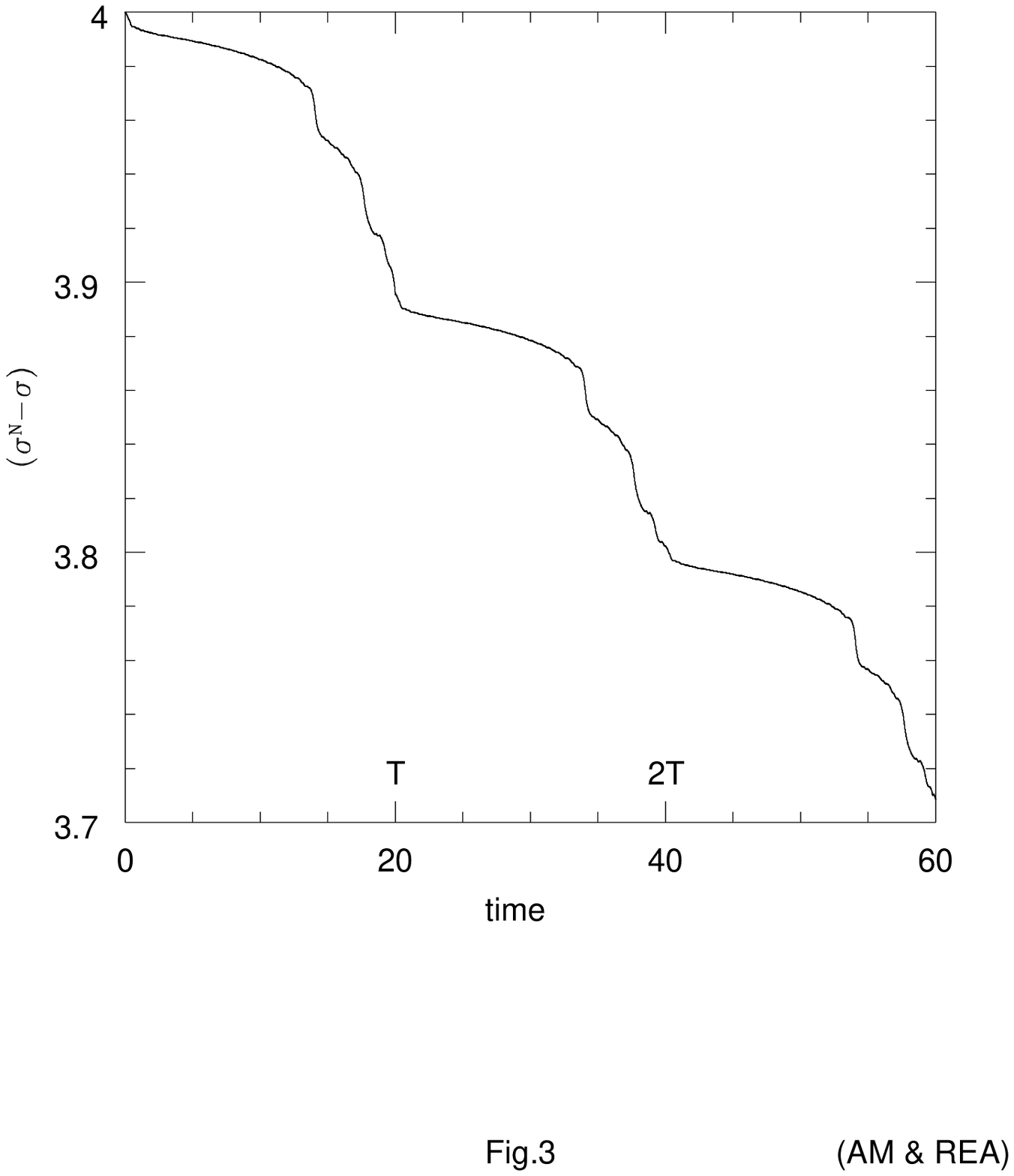}
\caption{ The plot shows the evolution of the difference
$\sigma'-\sigma$ as a function of time in the Lorenz system
(Eq.~(\ref{LORR})) with unknown parameter $\sigma$ when the given time
series of $x$ is truncated after the time $T=20$. We have used The finite
time series thrice and plotted the curve for the interval $0\leq t \leq
3T$. We see that the successive values of the difference at $t=0,T,2T,3T$
decrease. This indicates that a repetitive use of the finite time series
can improve the accuracy of parameter estimation.} \end{figure}

\begin{figure} 
\epsffile{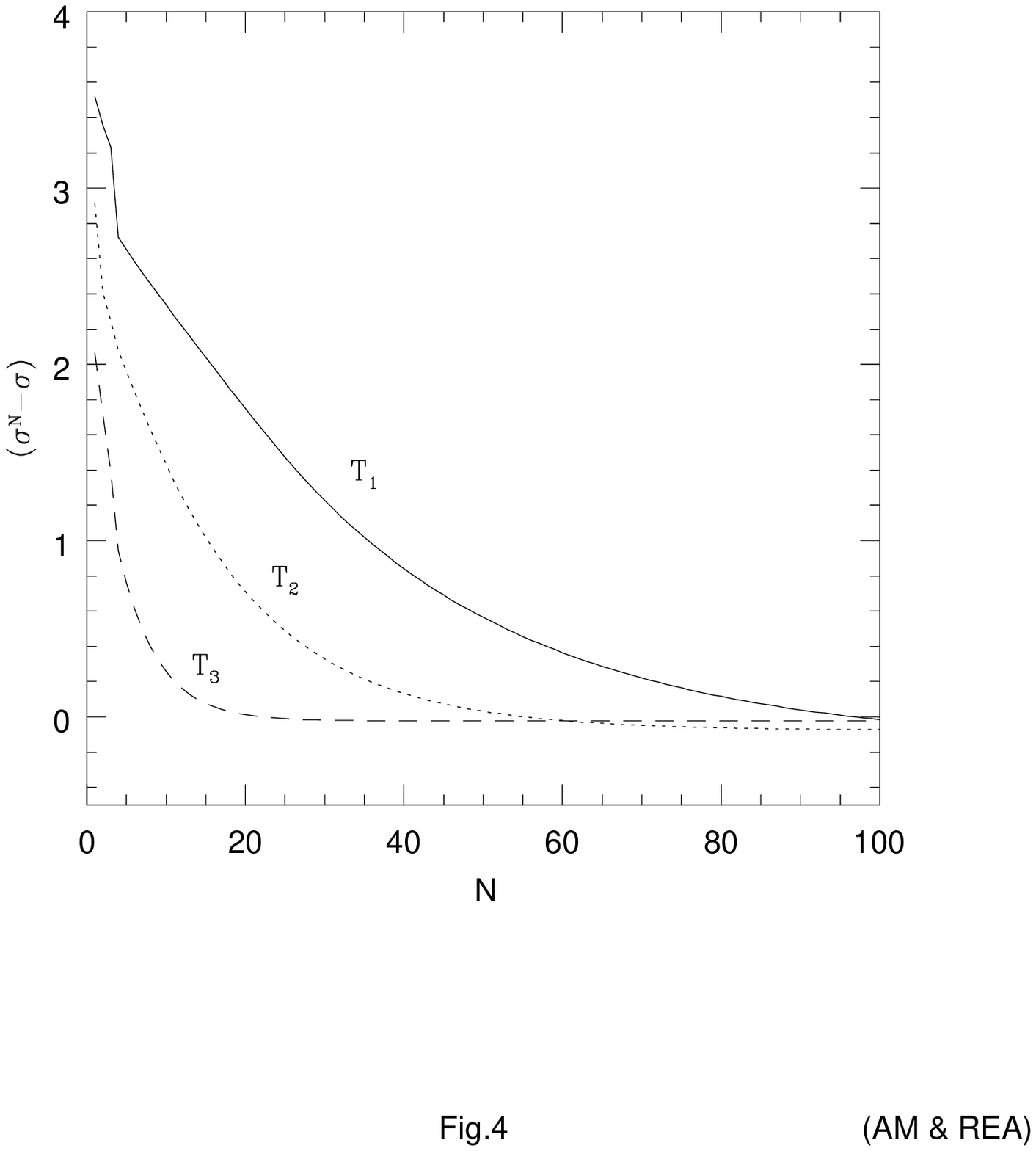} 
\caption{ The graph shows the successive differences
$\sigma^N-\sigma$ plotted as a function of $N$, the number of times a
finite time series $\{x(t)|o\leq t \leq T\}$ is used to estimate an
unknown $\sigma$ in a Lorenz system (Eq.~(\ref{LORR})). We see that after
an initial transient, the difference decreases showing better accuracy of
the estimation.  We also see that as $N$ increases further the accuracy of
estimation saturates and it is not possible to improve upon the estimation
beyond this. The three curves correspond to three different values of T
where $T_1<T_2<T_3$. It can be seen that a larger $T$ leads to a better
estimation as expected.} \end{figure}

\begin{figure}
\epsffile{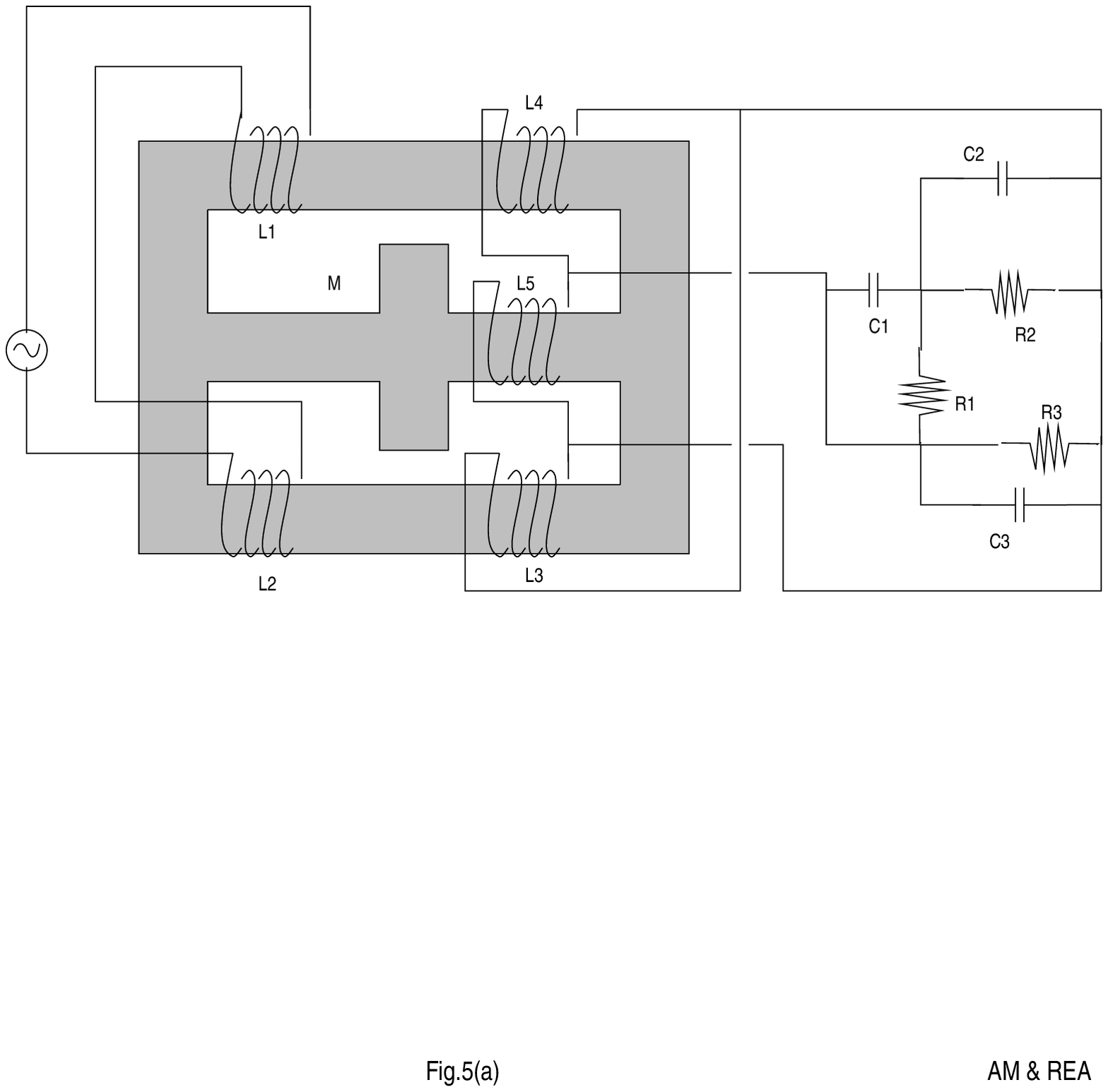}  
\epsffile{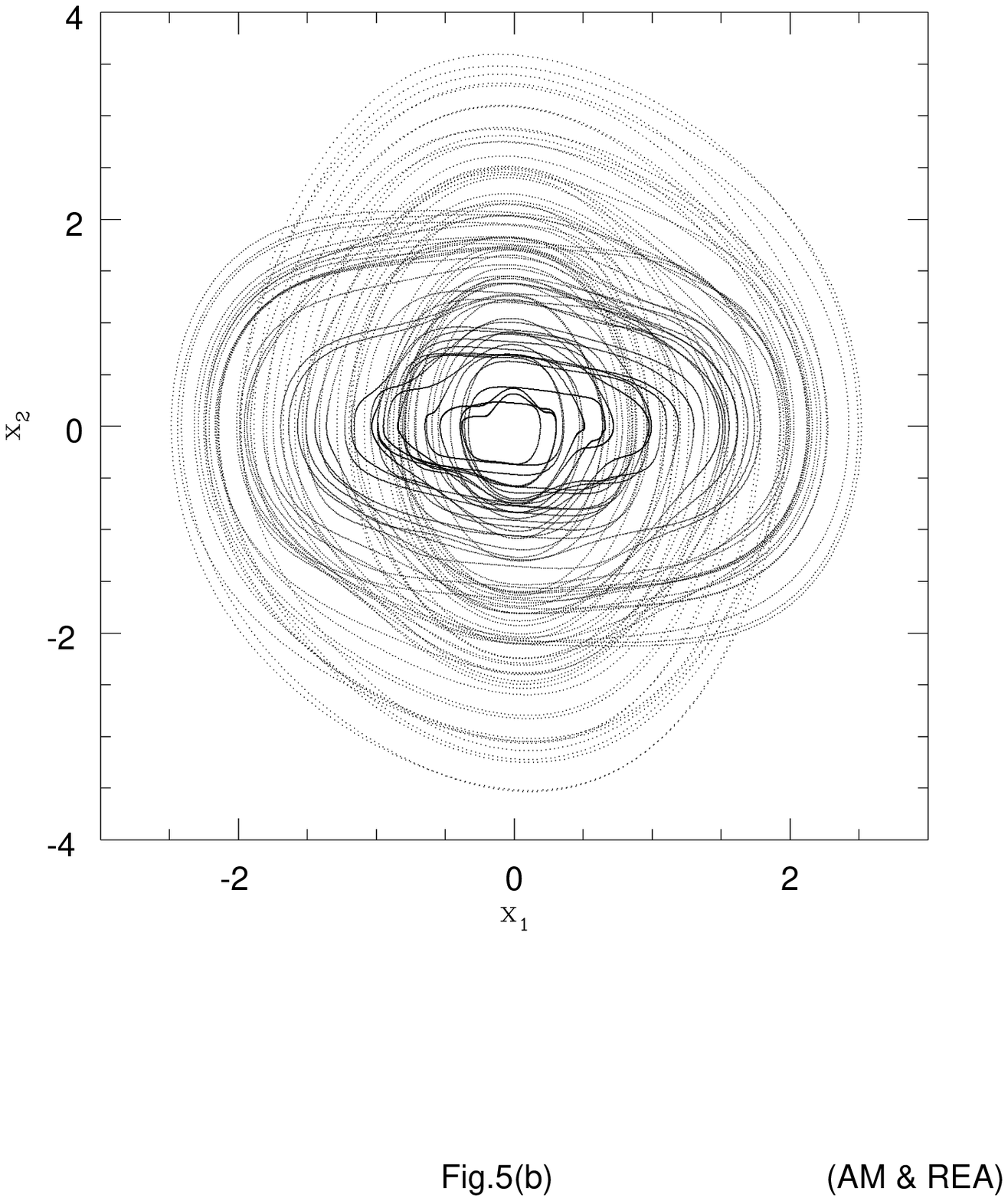}  
\caption{ A schematic diagram (Fig.~5(a)) of a phase
converter circuit (Eq.~(\ref{CKT})) which shows a chaotic behaviour.
Fig.~5(b) shows a chaotic attractor for the parameter values
$k=0.1,B=3.0$.} \end{figure}

\begin{figure} 
\epsffile{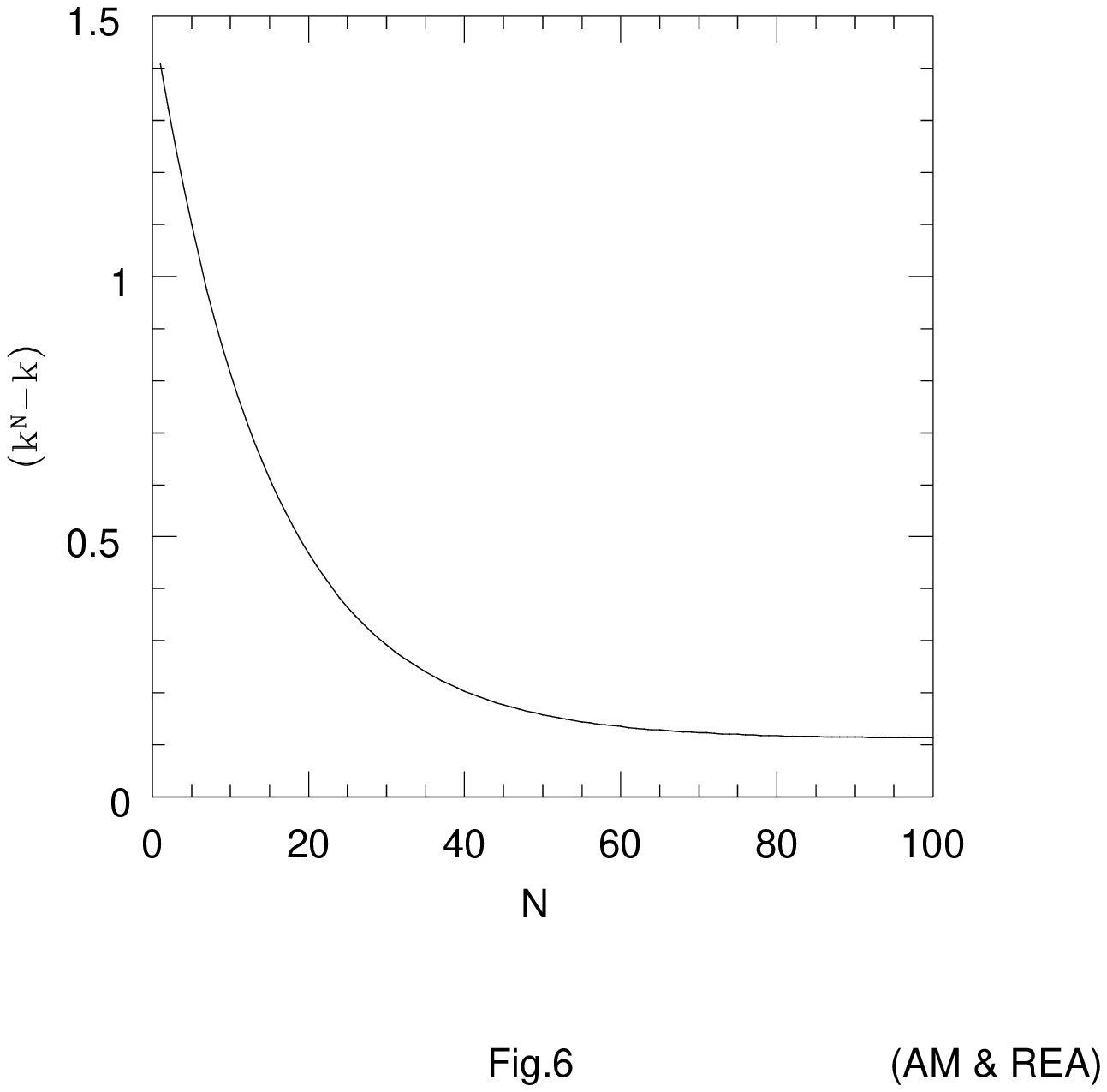} 
\caption{ The graph shows the successive differences
$k^N-k$ plotted as a function of $N$, the number of times a finite time
series $\{x_2(t)|o \leq t \leq T\}$ is used to estimate an unknown $k$ in
a phase converter circuit system (Eq.~(\ref{CKT})). We see that after an
initial transient, the difference decreases showing better accuracy of the
estimation. We also see that as $N$ increases further the accuracy of
estimation saturates and it is not possible to improve upon the estimation
beyond this using our method.} \end{figure}

\begin{figure} 
\epsffile{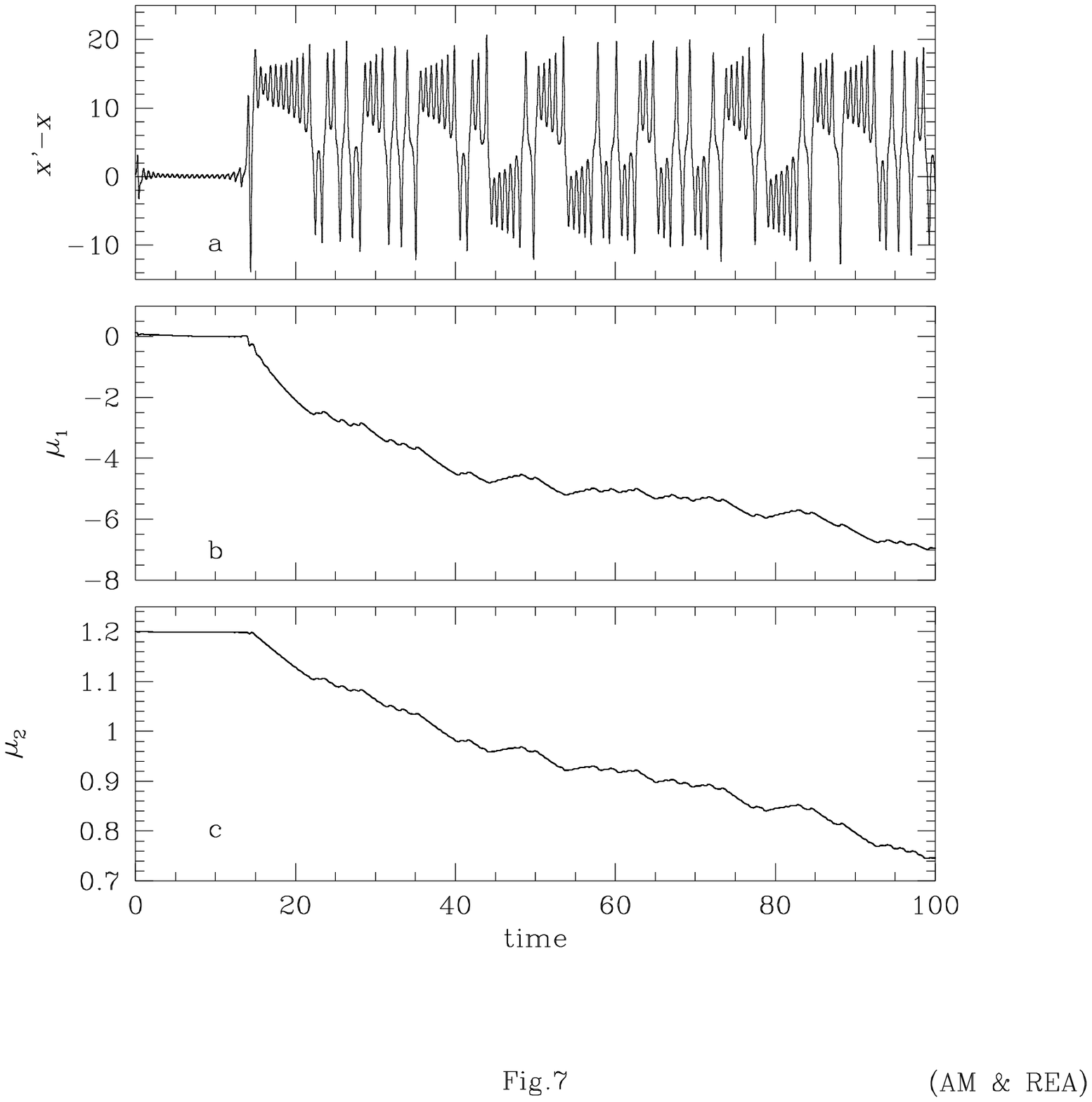} 
\caption{ The plots (a)-(c) show the time evolution of
$x'-x, \mu_1$ and $\mu_2$ respectively for the Lorenz system with the
feedback given in equation for $x$ and with the trial perturbation
function $G=\mu_1 x^2+\mu_2$ while the correct perturbation is
$F=A\sin(\omega x)$ (Eq.~(\ref{MODEL})). We see that the guess function
$G=\mu_1(Eq.~(\ref{MODEL})) x^2+\mu_2$ fails to produce synchronization
and hence can be discarded as a plausible model for $F$. It can also be
seen that there is no convergence of the parameters taking place.}
\end{figure}

\begin{figure}
\epsffile{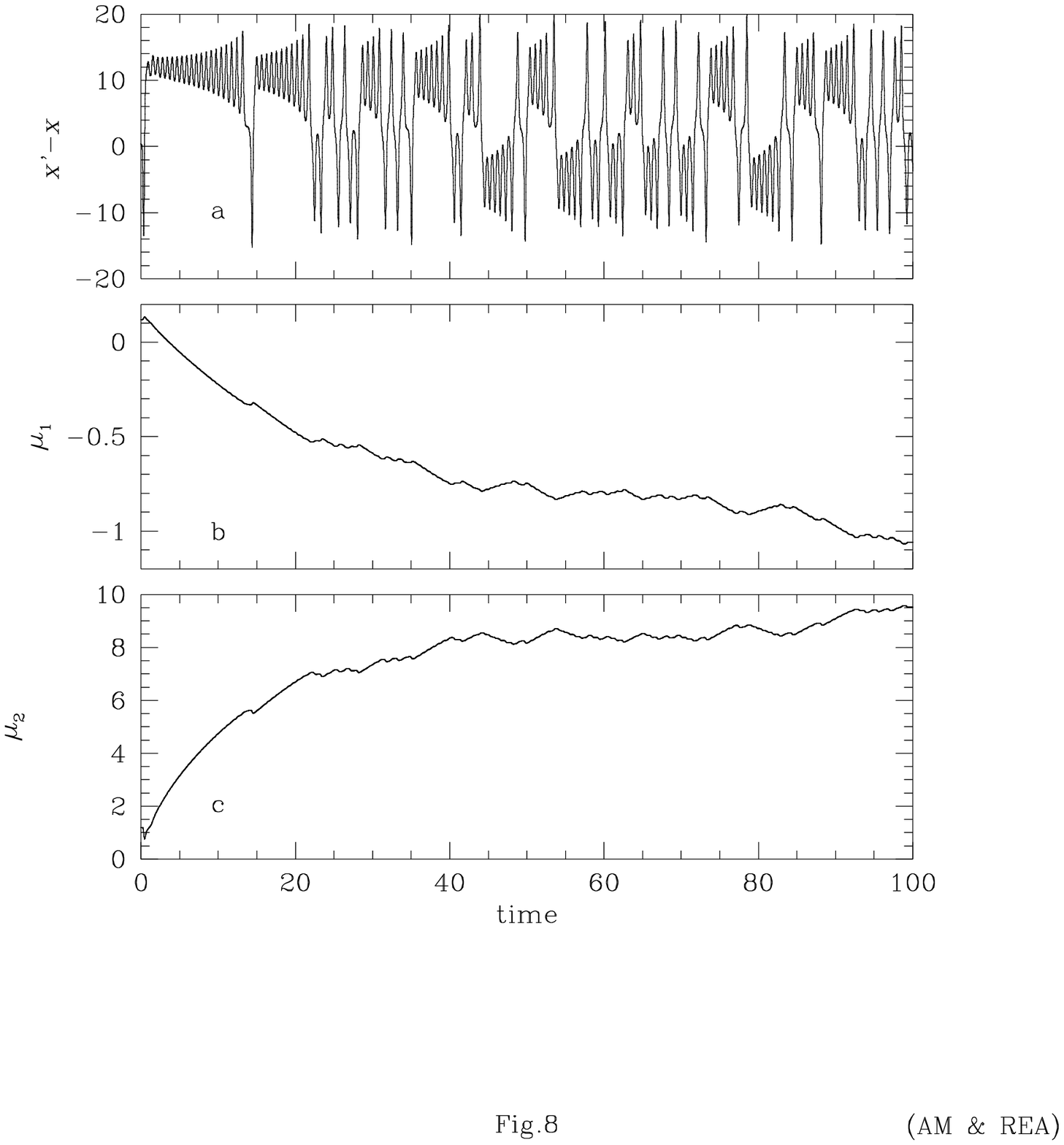}  
\caption{ The plots(a)-(c) show the time evolution of
$x'-x, \mu_1$ and $\mu_2$ respectively for the Lorenz system with the
feedback given in equation for $x$ and with the trial perturbation
function $G=\mu_1 x+\mu_2 x^3$ while the correct perturbation is
$F=A\sin(\omega x)$ (Eq.~(\ref{MODEL})). It can be clearly seen that even
when $G=\mu_1 x - \mu_2 x^3$ matches in form with $F$ upto two leading
terms in the expansion of $F$, it fails to produce synchronization and
hence can be discarded as a plausible model for $F$. Also there is no
convergence of the parameters taking place.} \end{figure}

\begin{figure} 
\epsffile{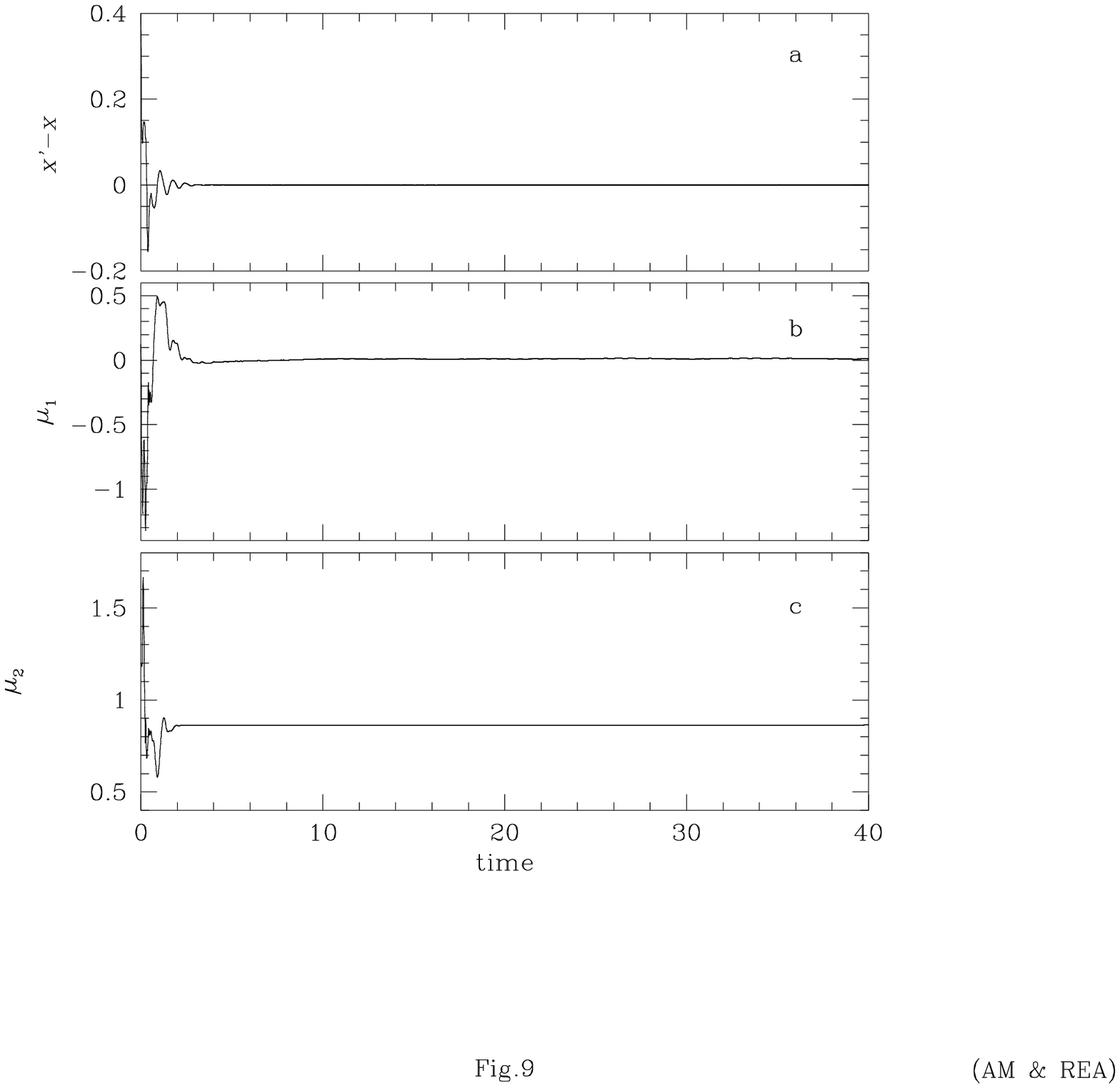} 
\caption{ The plots (a)-(c) show the time evolution of
$x'-x, \mu_1$ and $\mu_2$ respectively for the Lorenz system with the
feedback given in equation for $x$ and with the trial perturbation
function $G=\mu_1 \sin(\mu_2 x)$ while the correct perturbation is
$F=A\sin(\omega x)$ (Eq.~(\ref{MODEL})). It can be clearly seen that the
difference $x'-x$ converges to zero asymptotically indicating an exact
synchronization between the variables. Thus by using our method the guess
for the model perturbation function can be easily justified.} \end{figure}

\begin{figure}
\epsffile{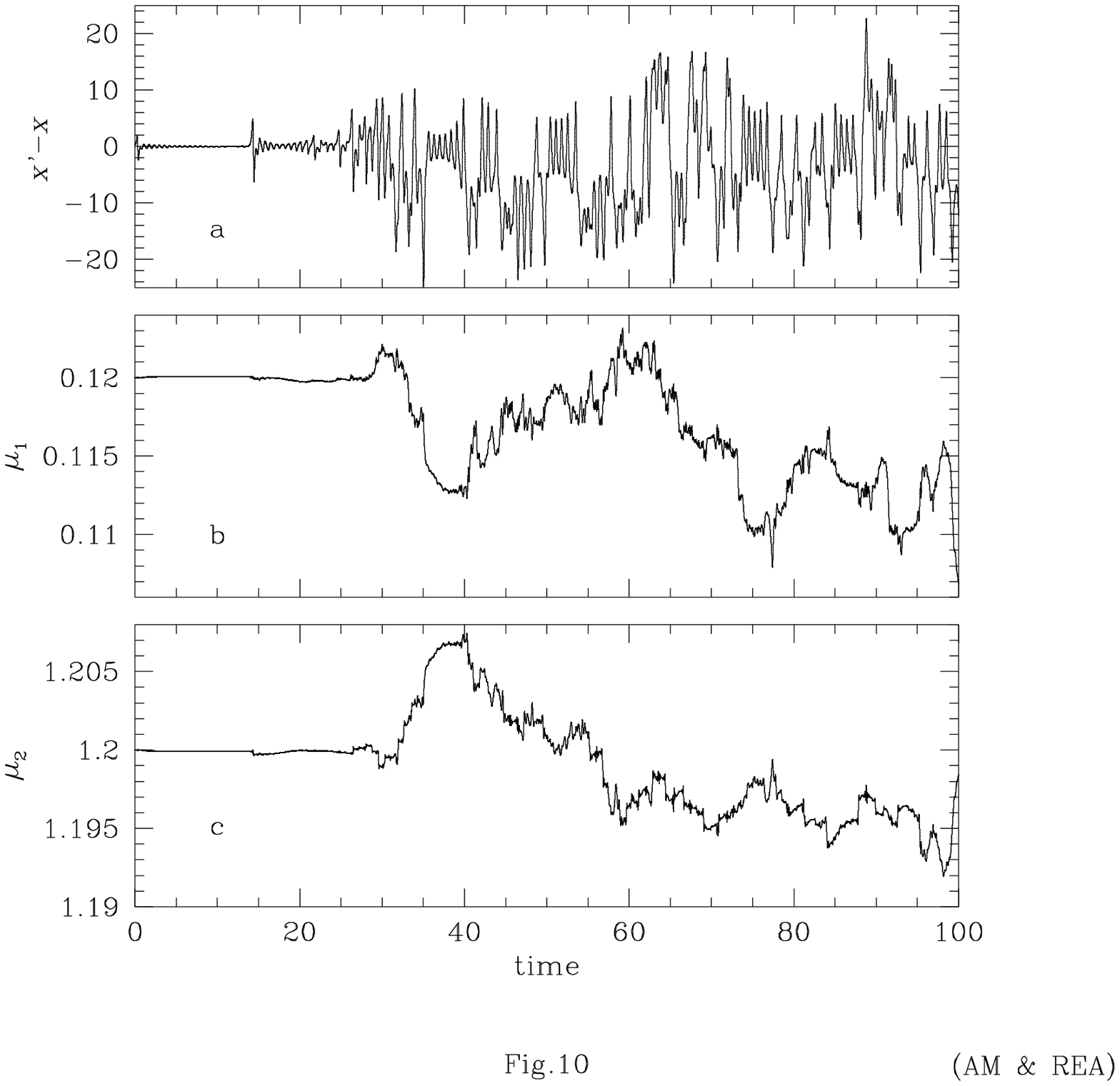} 
\caption{ The plots (a)-(c) show the time evolution of the
difference $x'-x$ and the parameters $\mu_1$ and $\mu_2$ respectively for
the Lorenz system with the feedback given in the equation for $x$ and with
the trial perturbation function $G=\mu_1 \sin(\mu_2 x)$ in the equation
for $y$ while the correct perturbation is $F=A\sin(\omega x)$ in the
equation for $x$ (Eq.~(\ref{MODELY})). Thus, unlike the case plotted in
Fig.9 the trial function used here, perturbs the wrong variable. It can be
clearly seen that the trial function $G$ does not produce synchronization
between variables. The parameters also do not converge.  Thus as expected,
the guess function $G=\mu_1(\sin\mu_2 x)$ when added to a wrong variable,
cannot model the perturbation.} \end{figure}


\begin{references}
\bibitem{Note} We note that the application of parameter estimation that
we have discussed here is fairly general and may not depend on a
particular method used to estimate the parameters. Use of alternative
methods for parameter estimation, such as those discussed in the
introduction may be possible.
\end{references}
\end{document}